%% file: main.tex
\documentclass[conference]{IEEEtran}
\usepackage{graphicx} % Required for inserting images
\usepackage{xcolor}
\usepackage{flushend}

\usepackage{booktabs}

\usepackage{listings}
\usepackage{paralist}
\usepackage{siunitx}
\usepackage{tikz}
\usepackage{tikz-uml}
\usepackage{pgfuml}
\usepackage{balance}
\usepackage{vcell}

\usepackage{hyperref}
\hypersetup{
    colorlinks=true,
    linkcolor=blue,
    filecolor=magenta,      
    urlcolor=cyan,
}

\usepackage{amsmath}
\usepackage{stmaryrd}

\definecolor{codegreen}{rgb}{0,0.6,0}
\definecolor{codegray}{rgb}{0.5,0.5,0.5}
\definecolor{codepurple}{rgb}{0.58,0,0.82}
\definecolor{backcolour}{rgb}{0.95,0.95,0.92}

\usepackage{amssymb}

\lstdefinelanguage{fml}{
  keywords={
    abstract,boolean,break,byte,case,catch,char,class,%
    const,continue,create,default,delete,do,double,else,extends,event,false,final,%
    finally,fire,float,for,goto,if,implements,import,instanceof,int,%
    interface,listen,log,long,multiverse,namespace,native,new,null,onfailure,private,protected,%
    public,return,short,static,super,switch,this,throw,%
    throws,transient,true,try,typedef,unique,use,void,values,volatile,while,with,
    as,begin,match,from,select,where,end,in},%
  morekeywords={
    [3]concept,model,forEach,assert%
  },%
  sensitive,%
  morecomment=[l]//,%
  morecomment=[s]{/*}{*/},%
  morestring=[b]",%
  morestring=[b]',%
}

\lstdefinestyle{mystyle}{
    backgroundcolor=\color{backcolour!40},
    commentstyle=\color{codegreen!60}\rmfamily,
    keywordstyle=\color{magenta},
    numberstyle=\tiny\color{codegray},
    stringstyle=\color{codepurple},
    basicstyle=\ttfamily\scriptsize,
    breakatwhitespace=false,
    breaklines=true,
    flexiblecolumns=true,
    captionpos=b,
    emptylines=0,
    tabsize=2,
    texcl=true
}

\newcommand{\smallsf}[1]{{\smaller\textsf{#1}}}

\DeclareTextFontCommand{\mytexttt}{\ttfamily\hyphenchar\font=45\relax}

\begin{document}
% \linenumbers

%\title{Modeling with the multiverse}
\title{Modeling in the Design Multiverse}
%\title{Multiverse modeling}
%\title{Modeling in the Multiverse}

\author{
\IEEEauthorblockN{Sylvain Guérin, Salvador Mart\'inez}
\IEEEauthorblockA{IMT Atlantique\\
Lab-STICC, UMR 6285\\
Brest, France\\
\{name.surname\}@imt-atlantique.fr}
\and
\IEEEauthorblockN{Ciprian Teodorov}
\IEEEauthorblockA{ENSTA | Institut Polytechnique de Paris\\
Lab-STICC, UMR 6285\\
Brest, France\\
\{ciprian.teodorov\}@ensta.fr}
}

%\author{\IEEEauthorblockN{Anonymous Authors}}

\maketitle

\begin{abstract}
Real-world design processes often involve the evolution and divergence of design paths (by branching, revising, merging, etc.), especially when multiple stakeholders or teams operate concurrently and/or explore different alternatives for complex and heterogeneous systems. Unfortunately, this variability in time and space can not be directly managed in current modeling spaces but requires resorting to external tools and methodologies. In order to tackle this problem, we introduce the \emph{Design Multiverse}. The \emph{Design Multiverse} aims to integrate in the modeling space a selection of revisions and variants, representing snapshots of a design state composed of multiple artifacts. This enables stakeholders to seamlessly trace, analyze, and manage design decisions, system variants, and their interdependencies. Concretely, in this paper we present a conceptual definition of the \emph{Design Multiverse}, discuss usage scenarios such as model product lines and model/metamodel co-evolution, and propose an implementation leveraging the model federation paradigm.

\end{abstract}

\section{Introduction}
\label{sec:intro}
\input{intro}

\section{Design Multiverse}
\label{sec:multivserse}
\input{multiverse}

\section{Usage scenarios}
\label{sec:scenarios}
\input{scenarios}

\section{Getting the Design Multiverse Through Model Federation}
\label{sec:approach}
\input{implem}

\section{Related work}
\label{sec:related}
\input{related}

\section{Conclusions \& future work}
\label{sec:conclusion}

\input{conclusion}

\bibliographystyle{ieeetr}
\bibliography{biblio}

\end{document}

%% file: intro.tex
%\medskip 

The INCOSE Systems Engineering Vision 2035\cite{vision2035} outlines a future of systems that are resilient, sustainable, and adaptable throughout their lifecycle. Yet, current engineering practices fall short of this ambition. Modern cyber-physical systems are increasingly heterogeneous, spanning diverse disciplines, tools, and models, which makes it difficult to ensure coherence and consistency across the design process. Additionally, the lack of formal mechanisms to track the \textbf{evolution of design decisions} results in disconnected variant histories, unclear rationales, and difficulty adapting to changing requirements. Traditional approaches often treat design decisions and variants as static and isolated entities, limiting responsiveness to change. Critically, the absence of a \textbf{temporal perspective on the design process} impedes understanding of how past choices shape present configurations and future behaviors, which is an essential capability for building robust and adaptable systems.

%The INCOSE Systems Engineering Vision 2035 \cite{vision2035} anticipates systems that are not only technically advanced but also resilient, sustainable, and adaptable over their lifecycle. Despite the ambition of this vision, several significant barriers impede the realization of its goals. One of the main challenges is the inability of current systems engineering practices to manage the sheer scale and complexity of modern cyber-physical systems. The increased \textbf{heterogeneity} across disciplines, tools, and models creates fragmentation, making it difficult to ensure coherence and consistency throughout all stages of the design and development process. Furthermore, there is a lack of formal mechanisms to track the \textbf{evolution of design decisions over time}, leading to disconnected records of system variants, unclear rationales behind design changes, and difficulties in integrating new components or adapting to evolving requirements. Indeed, traditional systems engineering practices are often ill-equipped to manage this increasing variability and tend to treat design decisions and system variants as static, isolated entities. This limits their ability to anticipate or respond to future changes and complexities. Furthermore, the lack of a \textbf{temporal perspective on the design process} makes it difficult to capture the relationships between past decisions, present configurations, and future system behaviors, which are crucial to ensure robust and adaptable system designs.

As a solution to the aforementioned issues, this paper introduces the \emph{Design Multiverse}. The design multiverse is a conceptual, problem-framing framework, which permits understanding and addressing current problems in system design approaches and methodologies which are related to the lack of time and space awareness (e.g., the co-evolution of related artifacts, the co-existence of revisions and variants, etc.) of current modeling spaces. Concretely, the \emph{Design Multiverse} integrates (e.g., makes co-exist) in the modeling space (e.g., model editors) a selection of revisions and variants, representing snapshots of a design state composed of multiple artifacts. In this setting, modelers are enabled to trace and analyze design decisions, system variants, and their inter-dependencies in a way that aligns with the future system engineering vision. 

We propose \emph{model federation}~\cite{amrani2024survey} as an infrastructure for the realization of the \emph{Design Multiverse}. Model federation is a multi-model management approach based on the use of virtual models and loosely coupled links. The models in a federation remain autonomous and are represented in their original technological spaces (and accessed through adapters), whereas virtual models (also called conceptual models) and links serve as control components used to present different views to the users and maintain synchronization through the use of automated behaviors. In the \emph{Design Multiverse} scenario, links are used to maintain references between artifacts, revisions, and variants, adapters are used to present to the user coherent and usable multi-version multivariant models (e.g., by the creation of variant types) and behaviors to support tasks such as migration and co-evolution.

The remainder of the paper is structured as follows. \autoref{sec:multivserse}  introduces the \emph{Design Multiverse}, followed by application scenarios in \autoref{sec:scenarios}. \autoref{sec:approach} presents a prototypical realization based on model federation. \autoref{sec:related} discusses related work. \autoref{sec:conclusion} concludes providing future directions.

%The remainder of the paper is organized as follows. Section~\ref{sec:multivserse} formally introduces the \emph{Design Multiverse} while the application scenarios are described in Section~\ref{sec:scenarios}. We present a prototype implementation of the \emph{Design Multiverse} leveraging model federation in Section~\ref{sec:approach}. Section~\ref{sec:related} discusses related work. We finish the paper in Section~\ref{sec:conclusion} by presenting conclusions and future work.

% \sm{In this paper we provide a formal definition of the \emph{Design Multiverse} and of the co-evolution operation and a prototype implementation leveraging model federation and Openflexo.}

% The idea here is to sell the integration of variability (in time and space) into the modeling space. We do not propose a new model of revisions and variants management~\cite{ananieva2019towards,berger_et_al:DagRep.9.5.1,linsbauer2017classification} in general, but an approach to integrate them in the modeling space.

% \medskip
% \textbf{Problems:}
% \begin{enumerate}
%     \item co-existence
%     \item conformance relation (e.g., typing relation between graphs)
%     \item migration
% \end{enumerate}

% \medskip
% \textbf{Objectives:}
% \begin{enumerate}
%     \item Integrate (a selection of) everything in the modeling space
%     \item Enable Multiverse manipulation / navigation
%     \item Deal with differences and complexity management
% \end{enumerate}

%% file: multiverse.tex
We introduce the \textbf{Design Multiverse} as a \emph{problem-framing framework} to understand and manage the complexities inherent in the evolution of design artifacts. Rather than assuming a linear trajectory of design activities, we acknowledge that real-world design processes often involve branching, merging, and divergence of design paths, especially when multiple stakeholders or teams operate concurrently, or when different alternatives are explored. The design multiverse captures these phenomena by modeling the design process as a \emph{partially ordered set of slices}, each representing a snapshot of a design state, connected through \textbf{design transitions}.

\begin{figure}
    % \hspace{-1cm}    
    \centering
    \includegraphics[width=0.47\textwidth]{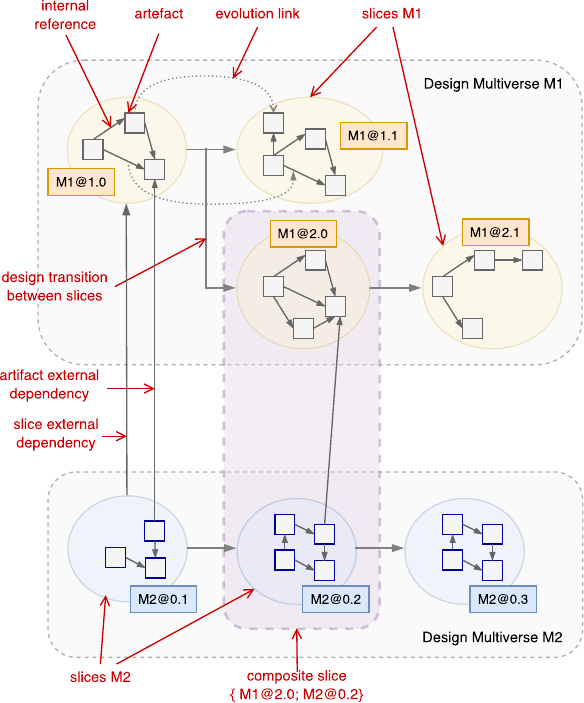}
    % essayer plutot abovecaptionskip
    % \vspace{-1cm}
    \caption{Key notions in the Design Multiverse conceptual framework.}
    \label{fig:design-multiverse}
 \end{figure}

The key notions are illustrated in figure \ref{fig:design-multiverse}, which presents two interlinked design multiverses $\mathcal{M}_1$ and $\mathcal{M}_2$. A \textbf{slice} is part of a design multiverse and is defined as a structured set of design \textbf{artifacts} (models, requirements, code elements, etc.), which are syntactic in nature. Tools and human interpreters operate over a slice to extract semantics, enabling reasoning, validation, and the formulation of design decisions that lead to the generation of new slices. The transitions between slices, capturing the local inter-slice design process, are deliberate evolutions that modify the design state based on rationale, analysis results, stakeholder input, or exploratory objectives.

Formally, a design multiverse is a directed acyclic graph, where nodes are slices and edges are \textbf{design transitions}. A slice is generally labeled with a version. From a given slice, multiple new slices may be created, forming branches that represent:
\begin{inparaenum}
\item Concurrent evolution by different teams;
\item Alternative responses to the same analysis results; or
\item Exploratory forks, where different hypotheses are evaluated in parallel.
\end{inparaenum} A slice is internally structured as rooted graph (with multiple roots potentially) where nodes are artifacts and edges \textbf{internal references} between artifacts (eg. conformance, usage, refinement, etc.). Design transition between slices can be accompanied by \textbf{evolution links}, which bind two successive versions of an artifact or an internal reference. Evolution links may carry semantic delta computation between the two versions (or a simple correspondence if artifact is unchanged), which eases coexistence and reasoning about multiple versions of a given artifact.

Working with multiple design multiverses implies \textbf{external dependencies}. External dependencies are defined between artifacts from different slices and design multiverse. These fine-grained dependencies may be aggregated and expressed as \textbf{slice external dependencies}.  

The usual workspace of a design process is a combination slices from the required design multiverses. We call this workspace a \textbf{composite slice}, illustrated in figure \ref{fig:design-multiverse} with version $2.0$ of $M1$ and version $0.2$ of $M2$. A composite slice is syntactically closed if all external dependencies are resolved inside the composite slice itself.

Together, these notions provide a structured foundation for modeling, analyzing, and reasoning about the inherently branching and evolving nature of real-world design processes.
%A design multiverse emphasizes  captures the essence of real-world engineering and system design processes. To illustrate the interest of using the \textbf{design multiverse} as a problem-framing framework, we describe in the following section a number of usage scenarios. 

%\cip{Ecrire un truc sur la navigation...}

%\paragraph{Navigating the multiverse in the modeling space}

%% file: scenarios.tex
In the following, we describe two categories of scenarios that benefit from the \emph{Design Multiverse} framing. Firstly, \emph{navigation scenarios}, which involve slices in the same multiverse, and secondly 
\emph{co-evolution scenarios}, which involve slices in at least two different multiverses.

\subsection{Navigating through revisions and variants}

The multiverse enables modeling experts to seamlessly navigate between revision slices and its contained elements through evolution links. This navigation occurs in the modeling space so that parts of the model would appertain to a certain revision while other parts may have been navigated forward or backward in time. This way of navigating revisions in the modeling space permits the evaluation of hypothesis. These hypotheses may come in the form of formal constraints (or properties) which are (re)-evaluated upon navigation and allow modelers to answer questions such as: Does this property hold in the current slice? Does it hold when navigating certain elements forward or backwards towards other slices?

\paragraph*{Product lines of modeling artifacts}
The product lines paradigm has been adapted to the modelware to capture model variants and provide an explicit way to manage them~\cite{haugen2004mda, heidenreich2008featuremapper, schwagerl2019integrated}. This has recently been extended to the management of adaptive modeling languages \cite{de2022modular, de2025adaptive}.

The design multiverse offer a natural representation for product line modeling: Each variant corresponds to a slice, and feature models can be viewed as abstractions over the multiverse structure. Merging, diffing and navigating design transitions (variants) can support variant management (e.g., migration operations) as well as reasoning.

\begin{table*}
\centering
\caption{Examples of Link-Scoped Co-evolution Across Design Multiverses}
\begin{tabular}{@{}lllll@{}}
\toprule
\textbf{Link Type} & \textbf{Source Multiverse} & \textbf{Target Multiverse} & \textbf{Co-evolution Trigger} & \textbf{Example} \\ \midrule
Conformance ($\chi$ relation) (\autoref{confo-co})        & Model                      & Metamodel                   & Metamodel evolution           & UML metamodel evolution \\
Use (\autoref{cots-co})              & Project                    & COTS Library                & COTS version change           & Log4j v1 $\rightarrow$ v2 \\
Implementation     & Design                     & Requirements                & Requirement evolution         & Safety-critical system requirements \\
Refinement         & Implementation                      & Formal Spec                 & Specification evolution       & Safety-critical system formalization \\
Binding            & PIM-PSM transformation                        & PM                         & PM evolution                 & New Java version, new RTOS version \\
\bottomrule
\end{tabular}
\label{tab:co-evolution-links}
\end{table*}

% \subsection{Co-evolution scenarios}

% \sm{here goes everything which is related with co-evolution.}

\subsection{Co-evolution in the Design Multiverse}
\label{sec:coev}
%\textcolor{blue}{
The concept of \textbf{design multiverse} becomes especially powerful when paired with the non-trivial operations encountered during complex systems modeling. Here we formally define and study \emph{co-evolution}: the process of maintaining the consistency of inter-multiverse links when a change occurs in one multiverse. This framing also serves as specification for the prototypical realization from \autoref{sec:approach}.
%This framing allows us to formally define and study \emph{co-evolution}: the process of maintaining the consistency of inter-slice or inter-multiverse links when a change occurs in one slice or multiverse.
%}

\begin{figure}
    \centering
    \includegraphics[width=0.8\linewidth]{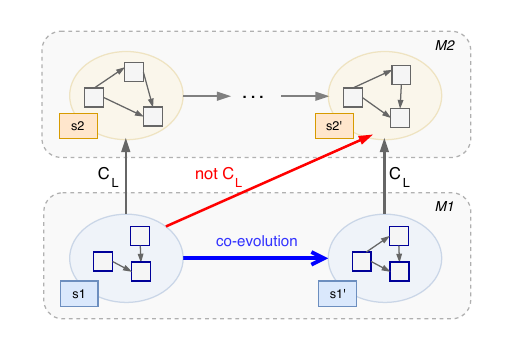}
    \caption{Co-evolution in the Design Multiverse}
    \label{fig:enter-label}
\end{figure}

Let $\mathcal{M}_1, \mathcal{M}_2$ be two design multiverses, and $s_1 \in \mathcal{M}_1$, $s_2 \in \mathcal{M}_2$ be slices. Now consider 
two arbitrary artifacts $a \in s_1$, $b \in s_2$ in these slices. If we have a relation $L(a, b)$ between these artifacts, 
we can define a \emph{semantic constraint} $\llbracket L(a, b) \rrbracket$ that captures the meaning of this relation (e.g., conformance, usage, refinement).

We define a consistency predicate between slices $s_1$ and $s_2$ as: \[
C_L(s_1, s_2) \iff \forall L(a, b), \llbracket L(a, b) \rrbracket \text{ holds}\]
% where $a \in s_1, b \in s_2\ \text{and }\llbracket L(a, b) \rrbracket$ holds.

In other words, all links of type $L$ between artifacts in $s_1$ and $s_2$ respect their semantic constraints.

If we assume a design transition (potentially transitive) in multiverse $\mathcal{M}_2$:
\[
\delta_2: s_2 \xrightarrow{d}_* s_2'
\]

Then, a \textbf{co-evolution} (under link type $L$) is a transition:
\[
\delta_1: s_1 \xrightarrow{co-ev} s_1' \quad \text{(in } \mathcal{M}_1)
\]
such that:
$
\neg C_L(s_1, s_2') \quad \text{and} \quad C_L(s_1', s_2') \text{ holds.}
$ That is, the evolution of $s_2$ to $s_2'$ breaks consistency, and a subsequent evolution of $s_1$ along $\delta_1$ restores it. Note that the co-evolution is triggered if and only if:
\[
C_L(s_1, s_2) \land \delta_2: s_2 \xrightarrow{d}_* s_2' \land \neg C_L(s_1, s_2')
\]
The co-evolution is central to maintaining consistency in distributed design activities, across model variants, revisions, and heterogeneous models. The \textbf{design multiverse} perspective provides a principled approach to reason about when and how generic co-evolution operations must be triggered. We can now consider different relation type between design multiverses to identify multiple co-evolution scenarios.

\subsubsection{Model/Metamodel Co-evolution}
\label{confo-co}

\begin{figure}
    \centering
    \includegraphics[width=1\linewidth]{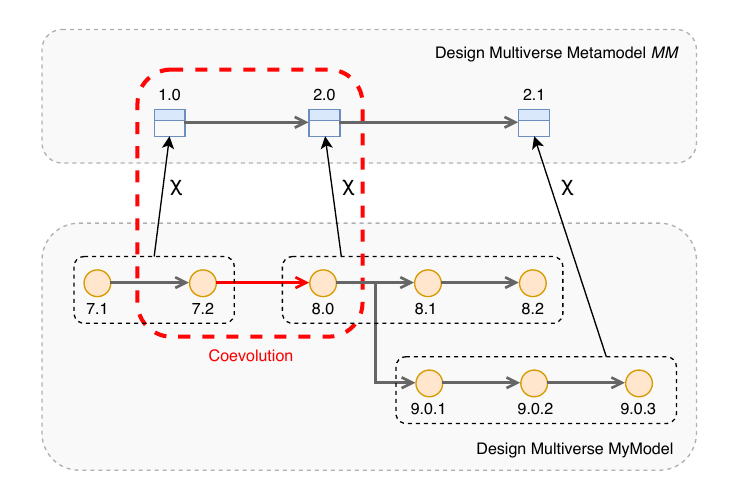}
    \caption{Concurrent evolutions of a model and its metamodel}
    \label{fig:evolution-history}
\end{figure}

\autoref{fig:evolution-history} illustrates the concurrent evolution of a model~\smallsf{MyModel} conforming to its metamodel \smallsf{Metamodel} ($\chi$ relationship). Both artifacts are designed and evolve in their own design multiverse: 
\[
\text{\smallsf{MyModel}}~\in~\mathcal{M}_1\text{, and \smallsf{Metamodel}}~\in~\mathcal{M}_2.
\]

Successive revisions of both metamodel and model, are slices in their respective \emph{design multiverse}. We rely on existing versioning systems (such as git, or a naming policy on the file system) to associate a label to a slice. In a typical design process and at a given time, the designer works with a version of the model and with a version of a metamodel in a composite slice (for example $\langle$ \smallsf{MyModel@(7.2)}, \smallsf{Metamodel@(1.0)} $\rangle$).

If \smallsf{Metamodel@(1.0)} evolves 
(for example, during the transition $\delta_2: \text{\smallsf{Metamodel@(1.0)}} \xrightarrow{d} \text{\smallsf{Metamodel@(2.0)}}$), 
and the conformance relation $\chi$ is not preserved, \smallsf{MyModel@(7.2)} must co-evolve to \smallsf{MyModel@(8.0)}, which is the classical case of model/metamodel co-evolution~\cite{cicchetti2008automating,hebig2016approaches}.

%At some points of the design multiverse graph, the conformance relationship $\chi$ is broken (for example, during the transition \smallsf{MyModel@(7.2)} $\rightarrow$ \smallsf{MyModel@(8.0)}) and requires a co-evolution scheme to adapt the model to the new version of the metamodel. 
Many approaches supporting operation-based versioning offer solutions to perform this transformation. This is made possible by the support of semantic deltas~\cite{modelsward13,van2013semantic}.  
However, some evolutions of a metamodel do not have any automatic solutions without changing the semantics or being ambiguous, and thus require manual decisions. In this situation the design multiverse offer the underlying structure needed to build a co-evolution specific modeling workspace (the red rounded rectangle in \autoref{fig:evolution-history}), which will allow the specification of the co-evolution transition $\delta_1$.

\subsubsection{COTS Evolution and Dependency Management}
\label{cots-co}
%When a project uses a Commercial Off-The-Shelf (COTS) component, a 'use' link connects the project's artifacts to the COTS artifacts. Each evolves in its own multiverse. If the COTS component is updated, the project must co-evolve to remain consistent with the new version.

Commercial Off-The-Shelf (COTS) components, such as libraries, are frequently integrated into software systems to reduce development time and leverage existing solutions. In the context of the design multiverse, the act of incorporating a COTS component creates a \textbf{usage link} between a slice in the software project’s multiverse and a slice in the COTS multiverse. When the COTS component evolves (e.g., due to a security update, performance improvement, etc.), the slice containing the COTS component moves forward in the COTS multiverse. This may break the semantic expectations or runtime behavior assumed by the project slice, thus triggering a need for co-evolution: the project must adapt, through new design decisions, to align with the updated COTS version while maintaining the validity of the usage link. %This could involve API refactoring, configuration changes, or dependency injection rework.

This scenario closely parallels the model/metamodel co-evolution case: both involve a directed semantic dependency—conformance in the model/metamodel case, and behavioral or structural dependence in the COTS case. In either case, evolution of the \emph{defining artifact} (metamodel or COTS) can invalidate assumptions in the dependent slice, triggering co-evolution to restore consistency. Tools that compare versions, detect incompatibilities, or suggest adaptations play similar roles in both contexts. However, notable differences exist. COTS components are typically black-boxes whose internal behavior may change across versions without syntactic cues, whereas metamodel evolution is usually explicit and syntactically interpretable. COTS evolution is often asynchronous, coarse-grained, and impacts many downstream systems, unlike the more controlled and synchronous evolution of metamodels within modeling environments. Moreover, COTS co-evolution introduces additional concerns, such as licensing, security, and supply chain constraints, that are generally absent in model/metamodel scenarios.

%This scenario closely parallels the model/metamodel co-evolution case. In both cases, there exists a directed semantic dependency, conformance in the model/metamodel case, and behavioral or structural dependence in the COTS case. In both, the evolution of the "defining" artifact (metamodel or COTS) may break assumptions made in the dependent slice, prompting the need for co-evolution to restore consistency. The usage of tools that can compare versions, highlight incompatibilities, or suggest adaptations plays a similar role to metamodel-aware co-evolution engines.

%However, there are also some differences. First, COTS components typically have a black-box character, their internal behavior may change across versions without syntactic cues visible to the client. In contrast, metamodel evolution is usually expressed within the modeling environment and can be interpreted syntactically. Second, the granularity and pace of evolution differ: COTS components may evolve independently and asynchronously, potentially affecting many dependent systems simultaneously, while metamodel evolution is often tightly controlled and co-evolves alongside its models within a modeling platform. Finally, licensing, security, and supply chain concerns add non-functional dimensions to the co-evolution of COTS components, which are not typically present in model/metamodel co-evolution.

Despite the prevalence of COTS components in software engineering, their role has been historically underrepresented in the modeling and Model-Driven Engineering (MDE) literature \cite{grandchallenges}. 
%Aside from simulation tools and a few platform integrations, COTS usage in the MDE ecosystem remains limited, lagging behind the practices of general software engineering. 
%If the modeling community had afforded COTS integration the same conceptual and tooling importance as metamodel-based conformance and model transformations, it is plausible that the adoption of COTS components within the modeling toolchain would be more widespread today. 
The design multiverse framework provides a unifying perspective that brings COTS usage and evolution into focus, not as an external engineering concern, but as an intrinsic part of the modeling space. 
%By treating each COTS version as a slice in its own multiverse, and its integration as a usage links into the project multiverse, the design multiverse highlights the need for rigorous co-evolution. 
Recognizing COTS evolution as a first-class co-evolution scenario through the multiverse lens could help promote more modular, extensible, and interoperable modeling environments.

Besides the conformance and usage scoped co-evolution, \autoref{tab:co-evolution-links} illustrates 3 more link-scoped co-evolution scenarios, induced by
\begin{inparaenum}
\item the implementation links, between a design and its requirements;
\item the refinement link between an implementation and its formal specification;
\item the binding link, between a PIM-PSM transformation (a transformation from fixed platform-independent-model (PIM) to a platform-dependent-model) and the evolving target platform model (PM). Note that in this case, in general, the design multiverse of the PIM-PSM transformation needs to be consistent with both the PIM and PM evolution in their respective design multiverses.
\end{inparaenum}

%\sm{citations needed here}
%\cite{schwagerl2019integrated}
%\cite{haugen2004mda}
%\cite{heidenreich2008featuremapper}
%Adaptive SPL based languages \cite{de2022modular, de2025adaptive}, %including automatic migration between variants.

%% file: implem.tex
While the Design Multiverse provides a powerful conceptual
lens for framing design variability, evolution, and co-evolution, 
this section shows that it can also be grounded (at least partially) in practical engineering tools. We present a prototypical realization based on the model federation paradigm, which enables the structured coexistence of revisions and variants within a unified modeling environment.

\subsection{Model federation}
\begin{figure}
    \centering
    \includegraphics[width=1.0\linewidth]{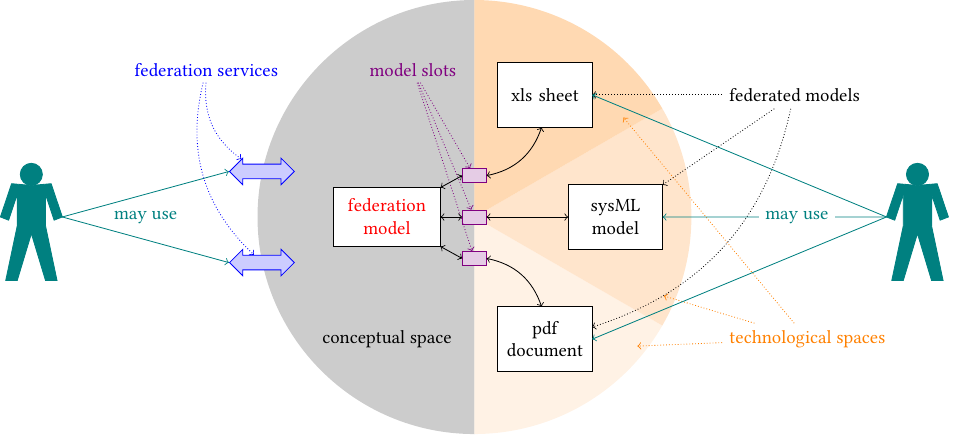}
    \caption{The principle of model federation}
    \label{fig:federation}
\end{figure}

Defined in ISO-14258~\cite{iso14258}, model federation introduces principles that complement the integration and unification approaches to interoperability. 
%The purpose of this standard is to characterize possible strategies to define correspondences between viewpoints, or languages associated with different engineering concerns. In this standard, 
The model federation approach, illustrated in \autoref{fig:federation}, keeps the models unchanged and independent. The semantic relations between models are encoded within a new element, the federation. 
%that holds the correspondences between the model elements. 
This allows the users to preserve their usual practice and tools, while an additional autonomous model, the 
\emph{federation model} is introduced to encode the dependencies, the interoperability semantics, and new integrated services. % illustrates the model federation approach.

%Many model federation approaches exist~\cite{amrani2024survey}, including Vitruvius\cite{Klare2021}, Epsilon\cite{Kolovos2008}, Reactive Links\cite{Ractiu2024}, comprehensive systems~\cite{stunkel2021comprehensive}, and OpenFlexo~\cite{bach202410}. 
%In this paper we rely on the latter framework, which offers the Flexo Modeling Language (FML) as a federation specific DSL to cover the needs of a federation model designer. 
%\emph{ModelSlot}s and libraries of \emph{Technological Adaptor}s provide proxys, that realize a bidirectional (read/write) connections to external information sources, giving them a contextual semantic/interpretation. 
%Models themselves are reified via \emph{VirtualModel}s comprising  of \emph{FlexoConcept}s interlinked with \emph{FlexoRole}s. When reified, the models and the links, can exhibit behaviors (\emph{FlexoBehavior}) and be reused. These constructs in FML allow the structured organization of the modeling workspace.

Many model federation approaches have been proposed~\cite{amrani2024survey}, including Vitruvius~\cite{Klare2021}, Epsilon~\cite{Kolovos2008}, Reactive Links~\cite{Ractiu2024}, comprehensive systems~\cite{stunkel2021comprehensive}, and OpenFlexo~\cite{bach202410}.
In this paper, we adopt the latter, which provides the Flexo Modeling Language (FML), a federation-specific DSL designed to support federation model development.
\emph{ModelSlot}s and libraries of \emph{Technological Adaptor}s serve as proxies, enabling bidirectional (read/write) connections to external sources and assigning them contextual semantics.
Models are reified through \emph{VirtualModel}s composed of \emph{FlexoConcept}s interlinked via \emph{FlexoRole}s.
Once reified, both models and links may exhibit behavior (\emph{FlexoBehavior}) and support reuse.
These constructs in FML facilitate a structured organization of the modeling workspace.

\subsection{Multiverses through federations}

\begin{figure}
    \centering
    \includegraphics[width=1\linewidth]{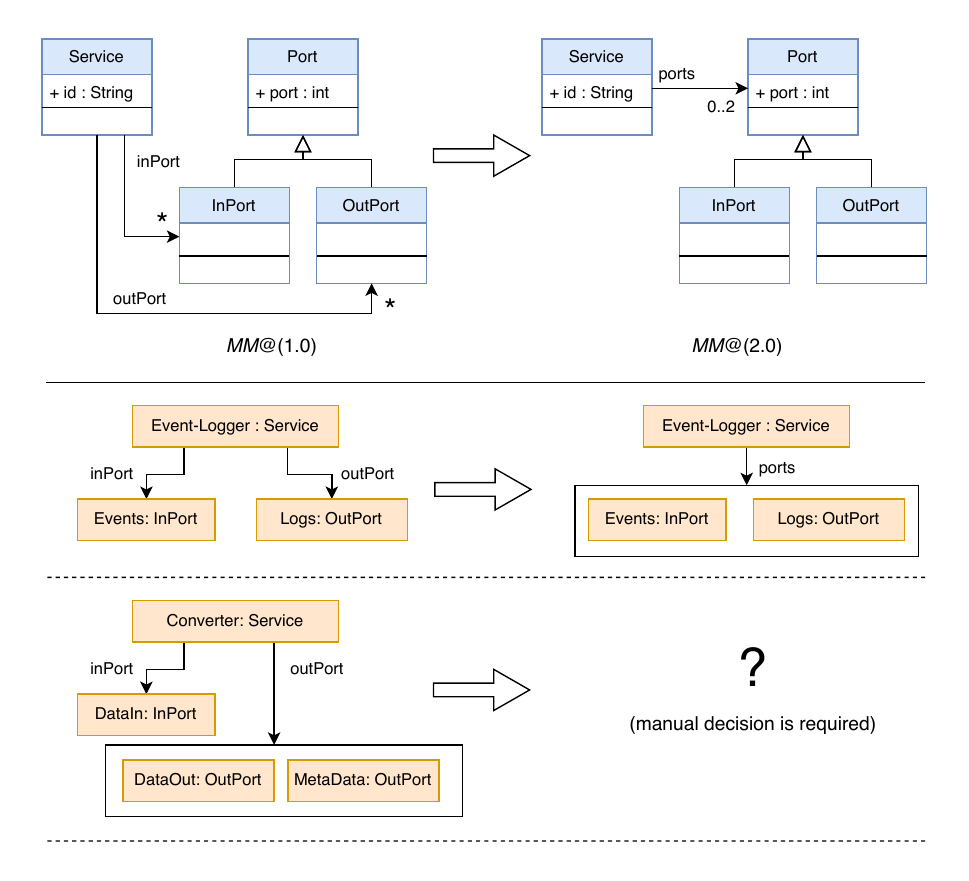}
    \caption{Examples of model/metamodel co-evolution}
    \label{fig:coevolution-example}
    %\sg{Mettre du texte à la place du point d'interrogation : co-evolution author has to decide manually the right domain-specific solution}
\end{figure}

In this section we illustrate the model/metamodel co-evolution scenario, given in \autoref{confo-co}, through an use case around a company hosting services for different customers, illustrated in Figure \ref{fig:coevolution-example}. 
In the use case, the metamodel is evolving, while some services are already defined. The automatic co-evolution of service \smallsf{EventLogger} is easily computable while the service \smallsf{Converter} requires a design decision. 
This use case was used in \cite{homolka2024don} as a motivating example for a co-evolution approach based on the co-existence of successive versions of the metamodel.
Instead of an integration approach, proposed in \cite{homolka2024don}, in this paper we propose a federation-based approach. The co-evolution workspace, illustrated in figure \ref{fig:evolution-history} by the red rounded rectangle, provides an environment where the four slices (the two metamodel versions and the two model versions) are accessible in the same modeling space as first-class citizen. 

%\sg{Ajouter la référence au papier NIER@ICSE pour parler de co-existence}: \cite{homolka2024don}

%\cip{TODO : Faire le lien avec la formalisation sur la co-évolution}

%\sm{Say... that we rely in existing versioning systems, that we only need tags and that we do need to modify those versions which remain in ther technological spaces.}

%\sm{Here we need to map the concepts of the multiverse to the federation. What are slices, what are composite slices, how to we create as federations?}

%\sg{parler de versioning and semantic deltas ?}

Listing \ref{listing:fmlcode-coevolution} shows an implementation of this co-evolution process in FML, through the definition of \smallsf{Coevolution} \emph{VirtualModel}. Metamodels are available in the environment through imports (lines 1 and 2) while the two versions of the co-evolved model are defined as \emph{ModelSlot} (lines 10 and 11). The co-evolution process is defined in behavior \smallsf{migrate()}, lines 14-29, which should be defined and externally called. Automated migration (based on the calculation of semantic deltas on evolution links) is performed on line 16, while the rest of the migration process handles the manual design decisions, which can be serialized in another artifact (\smallsf{portsBeingKept}) through model federation.

%\begin{figure}
\begin{lstlisting}[caption={FML code encoding co-evolution.},label={listing:fmlcode-coevolution}]
import EMFMetaModel MM_v1 from ["http://myMetaModel"@(1.0)];
import EMFMetaModel MM_v2 from ["http://myMetaModel"@(2.0)];
import EMFClass SERVICE_v1 from [MMv1:"Service"];
import EMFClass SERVICE_v2 from [MMv2:"Service"];
...
typedef EMFObjectType(eClass=SERVICE_v1) as ServiceV1;
typedef EMFObjectType(eClass=SERVICE_v2) as ServiceV2;
...
public model Coevolution
    EMFModel modelV1 with ModelSlot(metaModel=MM_v1);
    EMFModel modelV2 with ModelSlot(metaModel=MM_v2);
    Port[0..2] portsBeingKept;

    EMFModel migrate()
        // Automatic migration based on semantic delta
        Migration<MyModel> migration = migrate modelV2 from modelV1;
		
        ServiceV1 serviceV1 = modelV1.rootElement; 		
        ServiceV2 serviceV2 = migration.migrated(serviceV1);
        for (InPort inPort : serviceV1.inPorts)
            if (keepThatPort(inPort))
                // Rely on Migration which links migrated elements
                serviceV2.addToPorts(
                    migration.migrated(inPort));
        for (OutPort outPort : serviceV1.outPorts)
            if (keepThatPort(outPort)) {
                serviceV2.addToPorts(
                    migration.migrated(outPort));            
        return modelV2;

    // Encodes the decision on which port is to be kept
    boolean keepThatPort(Port port)
        return portsBeingKept.contains(port);
  \end{lstlisting}
%\caption{FML code encoding co-evolution}
%\label{fig:fmlcode-coevolution}
%\end{figure}

\subsection{Type System as an extension of MF}

Although the model federation approach is reasonable from a conceptual point of view, it must handle two type systems (eg. types \smallsf{ServiceV1} and \smallsf{ServiceV2}). This is cumbersome for the engineers, because an evolution often impacts a small part of the model. I would be more practical and comfortable to deal generically with everything that doesn't change during model co-evolution transition ($\delta_1$ in \autoref{sec:coev}).

To address this concern, we have introduced the notion of \textbf{partial design multiverse}, as opposed to \textbf{total design multiverse}, which exhaustively stores the entire design history. A partial design multiverse is defined from the explicit definition of slice labels (versions) by the computation of the minimal spanning subtree of the multiverse graph containing those explicit slices. Based on figure \ref{fig:evolution-history} partial design multiverse of metamodel $M\!M$ may be defined as :

{\footnotesize
\[
%\smallsf{multiverse}_{M\!M}[1.0] = \{ M\!M@1.0 \} \] \[
\smallsf{multiverse}_{M\!M}[1.0,2.0] = \{ M\!M@1.0, M\!M@2.0 \} \] \[
\smallsf{multiverse}_{M\!M}[1.0,2.1] = \{ M\!M@1.0, M\!M@2.0, M\!M@2.1 \} %\] \[
%\smallsf{multiverse}_{MyModel}[8.0,8.1,9.0.2] = \{ MyModel@8.0, MyModel@8.1, MyModel@9.0.1, MyModel@9.0.2 \}
\]
}

\autoref{listing:fmlcode-multiverse} illustrates the use of the \emph{design multiverse} for model federation in FML, which offers a support for multiverse nature of different elements. {\footnotesize$\smallsf{multiverse}_{M\!M}[1.0,2.0]$} is defined at line 1 as a partial multiverse embedding metamodels $M\!M@1.0$ and $M\!M@2.0$ presented in Figure~\ref{fig:coevolution-example}. \smallsf{SERVICE} class is defined at line 3 as a multiversed object from this multiverse metamodel. \smallsf{MyModel} and \smallsf{Service} also exhibit their multiverse nature (line 5 and 6). Assuming we take advantage of operation based versioning on evolution links, we introduce here the use of a multiverse-specific type system, based on the computation of semantic deltas between slices. 

\begin{lstlisting}[caption={FML code encoding co-evolution in multiverse context},label={listing:fmlcode-multiverse}]
import multiverse EMFMetaModel MM 
    from ["http://myMetaModel"]{@(1.0),@(2.0)}; 
import multiverse EMFClass SERVICE from [MM:"Service"];
...
typedef multiverse EMFModelType(eMetaModel=MM) as MyModel;
typedef multiverse EMFObjectType(eClass=SERVICE) as Service;
...
public model Coevolution
    // MyModel@(1.0) and MyModel@(2.0) are sub-types of MyModel 
    MyModel@(1.0) modelV1 with ModelSlot(metaModel=MM@(1.0));
    MyModel@(2.0) modelV2 with ModelSlot(metaModel=MM@(2.0));
    ...
    MyModel@(2.0) migrate()
        Migration<MyModel> migration = 
            migrate modelV2 from modelV1;

        // Service@(1.0) and Service@(2.0) are sub-types of Service 
        // Service could safely be used everywhere it does not need specialization
        Service@(1.0) serviceV1 = modelV1.rootElement; 		
        Service@(2.0) serviceV2
            = migration.migrated(serviceV1);
        for (InPort inPort : serviceV1.inPorts) {...}
        for (OutPort outPort : serviceV1.outPorts) {...}
        return modelV2;
    ...
\end{lstlisting}

\begin{figure}
    \centering
    \includegraphics[width=1\linewidth]{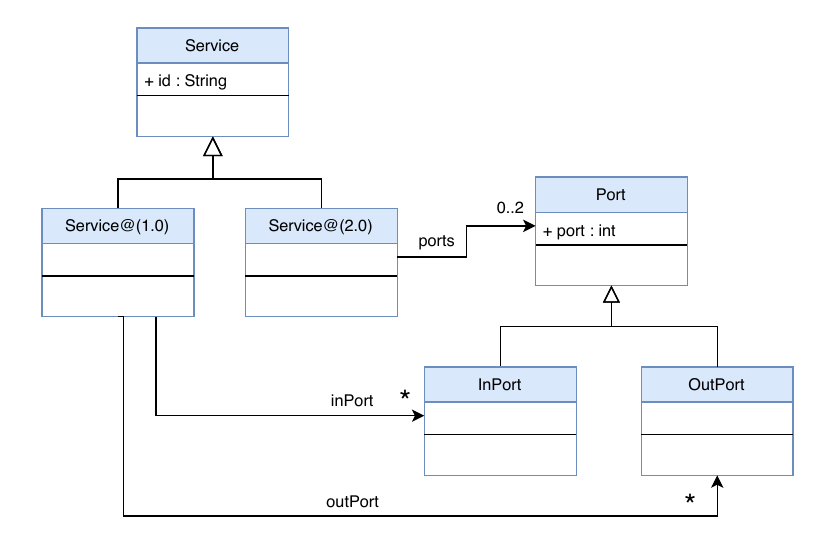}
    \caption{Handling multiverse with a dedicated type system}
    \label{fig:type-system}
 \end{figure}

Figure~\ref{fig:type-system} shows the type system being infered and used throughout \autoref{listing:fmlcode-multiverse}. \smallsf{Port}, \smallsf{InPort}, \smallsf{OutPort} remain unchanged accross {\footnotesize$\smallsf{multiverse}_{M\!M}[1.0,2.0]$} and can safely and generically be used by the engineer in that scope. On the other hand, \smallsf{Service} can generically be used but need a specialization \smallsf{Service@(1.0)} to deal with \smallsf{inPort}/\smallsf{outPort} features and \smallsf{Service@(2.0)} to deal with \smallsf{ports} features. One advantage of this approach is that the engineer can discover the inconsistencies by modifying the explicit versions of the partial multiverse, and interpreting the FML type errors.

% \begin{figure}

%
% \end{figure}

The model federation paradigm appears to be an ideal fit for a design multiverse infrastructure. The loosely coupled links allow interlinking slice elements from different design multiverses, grouping them at will to ensure their co-existence in the modeling space. Adapters preserve existing models in their original technological spaces and offer convenient access to their contents. Finally, behaviors capture the design transitions such as the migration from one revision to another.

%% file: related.tex
The design multiverse implies the integration of time (revisions) and space (variants) into the modelware. These two dimensions have attracted the attention of the modeling community from different points of view.

From a static perspective, as infrastructure to store and manage model histories including features such as conflict detection and merging, many works propose solutions to adapt the classical text-based versioning to the modeling realm~\cite{altmanninger2009survey, taentzer2014fundamental, sharbaf2023conflict}. From a software product line (SPL) perspective, in~\cite{schwagerl2019integrated} the authors present \emph{SuperMOD}, which integrates revision and variation control of evolving model-driven software product lines (here the artefacts of the product are models). Similarly, in~\cite{ananieva2022preserving} the authors present a SPL based approach which deals with revisions and additionally with the consistency of the models that form part of the different products that can be derived from the SPL. None of these works focuses on the co-existence of revisions and variants in the modeling space but only in the repository space. In~\cite{schropfer2022projectional} the authors present a text-based projectional editor for domain models which is variant aware, but does not integrate revisions.

From a runtime perspective, Temporal UML~\cite{cabot2003representing} proposes an UML profile that integrates temporal information into models so that it is possible to perform temporal (OCL) queries on runtime instances. Temporal EMF~\cite{gomez2018temporalemf}  adapts and extends this profile to EMF models, including an efficient database persistence layer. In~\cite{hartmann2014native} the authors present an approach (implemented in KMF~\cite{francois2014kevoree}) to enable the versioning of model elements including operations to navigate history at the application level. Datatime~\cite{lyan2021datatime, lyan2023reasoning} is an extension of models@runtime that integrates time (as in time series) and space (as changes in a graph topology) variability so that it is possible to reason about past, present and future states of the system. GreyCat\cite{hartmann2019greycat} presents many-worlds graphs. It focuses on instance models, that can not only be versioned, but forked in different worlds that will then follow different histories. All these approaches focus on runtime models instead of design (as we do in this paper), and do not deal with (or do it only marginally) the relationship of multiple interrelated evolving models. More similar to our approach but focused on executable models and not dealing with co-existing revisions and variants, \emph{Live modeling}~\cite{van2019multi,bagherzadeh2019live,bagherzadeh2021live,exelmans2024operation} proposes support to reflect model changes on runtime instances.

The multiverse have been also explored in the debugging community, which has coined the term multiverse debugging to refer to the possibility not only to go back in time to analyze past states, but also to branch to different possible future~\cite{lopez2019multiverse,pasquier2022practical,pasquier2023debugging}. From a conceptual point of view, the \emph{Digital Multiverse} has been proposed with the objective of framing the existence of variants (which may correspond to alternative designs, configurations, usage purposes) for Digital twins~\cite{cimino2021harmonising}. However, here again the focus is not design-phase modeling nor dealing with the co-existence problem from a technical point of view.

Besides the space and time integration of variability, the \emph{Design Multiverse} presented here identifies the need of handling the evolution and variability locally on each design space, and orthogonally using link-scoped conformance for reasoning about the binding between design multiverse. The numerous research effort around model/metamodel co-evolution \cite{cicchetti2008automating,hebig2016approaches} are founded on this dichotomy, but currently fall short to generalize, beyond type conformance, to the multitude of semantic constraints imposed by the other relations discussed in \autoref{tab:co-evolution-links}.

%% file: conclusion.tex
%In this paper, we have presented the \emph{design multiverse}, a framework for the reification and integration of the inherent variability of real-world design processes (this, is, revisions and variants) into the modeling space. We have discussed possible usage scenarios and proposed an initial implementation leveraging the model federation paradigm.

This paper introduced the \emph{Design Multiverse}, a framework for reifying and integrating the inherent variability of real-world design processes, namely revisions and variants, into the modeling space. We discussed several usage scenarios and proposed an initial implementation based on the model federation paradigm.

As future work, we plan to extend our experimentation to industrial-scale case studies and explore broader forms of co-evolution (beyond model/metamodel). Furthermore, we plan to investigate how design methodologies, currently informal guides through the space of design choices, can be reinterpreted as formal constraints over the multiverse evolution. Making their structure explicit within the \emph{Design Multiverse}, could enable formal reasoning about their intent, support their comparison and validation, and lay the groundwork for tool-assisted and semantically grounded methodology execution.

%% file: main.bbl
\begin{thebibliography}{10}

\bibitem{vision2035}
INCOSE, ``{Systems Engineering Vision 2035. Engineering Solutions for a Better
  World},'' 2022.

\bibitem{amrani2024survey}
M.~Amrani, R.~Mittal, M.~Goul{\~a}o, V.~Amaral, S.~Gu{\'e}rin,
  S.~Mart{\'\i}nez, D.~Blouin, A.~Bhobe, and Y.~Hallak, ``A survey of
  federative approaches for model management in mbse,'' in {\em Proceedings of
  the ACM/IEEE 27th International Conference on Model Driven Engineering
  Languages and Systems}, pp.~990--999, 2024.

\bibitem{haugen2004mda}
{\O}.~Haugen, B.~M{\o}ller-Pedersen, J.~Oldevik, and A.~Solberg, ``An
  mda{\textregistered}-based framework for model-driven product derivation,''
  {\em Software Engineering and Applications, USA}, 2004.

\bibitem{heidenreich2008featuremapper}
F.~Heidenreich, J.~Kopcsek, and C.~Wende, ``Featuremapper: mapping features to
  models,'' in {\em Companion of the 30th international conference on Software
  engineering}, pp.~943--944, 2008.

\bibitem{schwagerl2019integrated}
F.~Schw{\"a}gerl and B.~Westfechtel, ``Integrated revision and variation
  control for evolving model-driven software product lines,'' {\em Software and
  Systems Modeling}, vol.~18, pp.~3373--3420, 2019.

\bibitem{de2022modular}
J.~de~Lara, E.~Guerra, and P.~Bottoni, ``Modular language product lines: a
  graph transformation approach,'' in {\em Proceedings of the 25th
  International Conference on Model Driven Engineering Languages and Systems},
  pp.~334--344, 2022.

\bibitem{de2025adaptive}
J.~de~Lara and E.~Guerra, ``Adaptive modelling languages: Abstract syntax and
  model migration,'' {\em ACM Transactions on Software Engineering and
  Methodology}, vol.~34, no.~3, pp.~1--54, 2025.

\bibitem{cicchetti2008automating}
A.~Cicchetti, D.~Di~Ruscio, R.~Eramo, and A.~Pierantonio, ``Automating
  co-evolution in model-driven engineering,'' in {\em 2008 12th International
  IEEE enterprise distributed object computing conference}, pp.~222--231, IEEE,
  2008.

\bibitem{hebig2016approaches}
R.~Hebig, D.~E. Khelladi, and R.~Bendraou, ``Approaches to co-evolution of
  metamodels and models: A survey,'' {\em IEEE Transactions on Software
  Engineering}, vol.~43, no.~5, pp.~396--414, 2016.

\bibitem{modelsward13}
M.~Stephan and J.~R. Cordy, ``A survey of model comparison approaches and
  applications,'' in {\em Proceedings of the 1st International Conference on
  Model-Driven Engineering and Software Development - Volume 1: MODELSWARD,},
  pp.~265--277, INSTICC, SciTePress, 2013.

\bibitem{van2013semantic}
T.~van Der~Storm, ``Semantic deltas for live dsl environments,'' in {\em 2013
  1st International Workshop on Live Programming (LIVE)}, pp.~35--38, IEEE,
  2013.

\bibitem{grandchallenges}
A.~Bucchiarone, J.~Cabot, R.~F. Paige, and A.~Pierantonio, ``Grand challenges
  in model-driven engineering: an analysis of the state of the research,'' {\em
  Software and Systems Modeling}, vol.~19, no.~1, pp.~5--13, 2020.

\bibitem{iso14258}
{{ISO}/{TC} 184/{SC} 5 Interoperability, integration, and architectures for
  enterprise systems and automation applications committee}, ``{Industrial
  automation systems and integration -- Concepts and rules for enterprise
  models},'' Standard ISO 14258:1998, International Organization for
  Standardization, Geneva, CH, 1998.

\bibitem{Klare2021}
H.~Klare, M.~E. Kramer, M.~Langhammer, D.~Werle, E.~Burger, and R.~Reussner,
  ``Enabling consistency in view-based system development — the vitruvius
  approach,'' {\em Journal of Systems and Software}, vol.~171, p.~110815, 2021.

\bibitem{Kolovos2008}
D.~Kolovos, R.~Paige, and F.~Polack, ``Detecting and repairing inconsistencies
  across heterogeneous models,'' in {\em International Conference on Software
  Testing, Verification, and Validation}, pp.~356--364, 2008.

\bibitem{Ractiu2024}
C.-C. Ra{\c{t}}iu, W.~K. Assun{\c{c}}{\~a}o, E.~Herac, R.~Haas, C.~Lauwerys,
  and A.~Egyed, ``Using reactive links to propagate changes across engineering
  models,'' {\em Software and Systems Modeling}, pp.~1--27, 2024.

\bibitem{stunkel2021comprehensive}
P.~St{\"u}nkel, H.~K{\"o}nig, Y.~Lamo, and A.~Rutle, ``Comprehensive systems: a
  formal foundation for multi-model consistency management,'' {\em Formal
  Aspects of Computing}, vol.~33, no.~6, pp.~1067--1114, 2021.

\bibitem{bach202410}
J.-C. Bach, A.~Beugnard, J.~Champeau, F.~Dagnat, S.~Gu{\'e}rin, and
  S.~Mart{\'\i}nez, ``10 years of model federation with openflexo: challenges
  and lessons learned,'' in {\em Proceedings of the ACM/IEEE 27th International
  Conference on Model Driven Engineering Languages and Systems}, pp.~25--36,
  2024.

\bibitem{homolka2024don}
M.~Homolka, L.~Marchezan, W.~K. Assun{\c{c}}{\~a}o, and A.~Egyed, ``" don’t
  touch my model!" towards managing model history and versions during metamodel
  evolution,'' in {\em Proceedings of the 2024 ACM/IEEE 44th International
  Conference on Software Engineering: New Ideas and Emerging Results},
  pp.~77--81, 2024.

\bibitem{altmanninger2009survey}
K.~Altmanninger, M.~Seidl, and M.~Wimmer, ``A survey on model versioning
  approaches,'' {\em International Journal of Web Information Systems}, vol.~5,
  no.~3, pp.~271--304, 2009.

\bibitem{taentzer2014fundamental}
G.~Taentzer, C.~Ermel, P.~Langer, and M.~Wimmer, ``A fundamental approach to
  model versioning based on graph modifications: from theory to
  implementation,'' {\em Software \& Systems Modeling}, vol.~13, no.~1,
  pp.~239--272, 2014.

\bibitem{sharbaf2023conflict}
M.~Sharbaf, B.~Zamani, and G.~Suny{\'e}, ``Conflict management techniques for
  model merging: a systematic mapping review,'' {\em Software and Systems
  Modeling}, vol.~22, no.~3, pp.~1031--1079, 2023.

\bibitem{ananieva2022preserving}
S.~Ananieva, T.~K{\"u}hn, and R.~Reussner, ``Preserving consistency of
  interrelated models during view-based evolution of variable systems,'' in
  {\em Proceedings of the 21st ACM SIGPLAN International Conference on
  Generative Programming: Concepts and Experiences}, pp.~148--163, 2022.

\bibitem{schropfer2022projectional}
J.~Schr{\"o}pfer, T.~Buchmann, and B.~Westfechtel, ``Projectional editing of
  software product lines using multi-variant model editors,'' {\em SN Computer
  Science}, vol.~4, no.~1, p.~35, 2022.

\bibitem{cabot2003representing}
J.~Cabot, A.~Oliv{\'e}, and E.~Teniente, ``Representing temporal information in
  uml,'' in {\em International Conference on the Unified Modeling Language},
  pp.~44--59, Springer, 2003.

\bibitem{gomez2018temporalemf}
A.~G{\'o}mez, J.~Cabot, and M.~Wimmer, ``Temporalemf: A temporal metamodeling
  framework,'' in {\em International Conference on Conceptual Modeling},
  pp.~365--381, Springer, 2018.

\bibitem{hartmann2014native}
T.~Hartmann, F.~Fouquet, G.~Nain, B.~Morin, J.~Klein, O.~Barais, and
  Y.~Le~Traon, ``A native versioning concept to support historized models at
  runtime,'' in {\em Model-Driven Engineering Languages and Systems: 17th
  International Conference, MODELS 2014, Valencia, Spain, September 28--October
  3, 2014. Proceedings 17}, pp.~252--268, Springer, 2014.

\bibitem{francois2014kevoree}
F.~Francois, G.~Nain, B.~Morin, E.~Daubert, O.~Barais, N.~Plouzeau, and J.-M.
  J{\'e}z{\'e}quel, ``Kevoree modeling framework (kmf): Efficient modeling
  techniques for runtime use,'' {\em arXiv preprint arXiv:1405.6817}, 2014.

\bibitem{lyan2021datatime}
G.~Lyan, J.-M. J{\'e}z{\'e}quel, D.~Gross-Amblard, and B.~Combemale,
  ``Datatime: a framework to smoothly integrate past, present and future into
  models,'' in {\em 2021 ACM/IEEE 24th International Conference on Model Driven
  Engineering Languages and Systems (MODELS)}, pp.~134--144, IEEE, 2021.

\bibitem{lyan2023reasoning}
G.~Lyan, J.-M. J{\'e}z{\'e}quel, D.~Gross-Amblard, R.~Lefeuvre, and
  B.~Combemale, ``Reasoning over time into models with datatime,'' {\em
  Software and Systems Modeling}, vol.~22, no.~5, pp.~1689--1712, 2023.

\bibitem{hartmann2019greycat}
T.~Hartmann, F.~Fouquet, A.~Moawad, R.~Rouvoy, and Y.~Le~Traon, ``Greycat:
  Efficient what-if analytics for data in motion at scale,'' {\em Information
  Systems}, vol.~83, pp.~101--117, 2019.

\bibitem{van2019multi}
Y.~Van~Tendeloo, S.~Van~Mierlo, and H.~Vangheluwe, ``A multi-paradigm modelling
  approach to live modelling,'' {\em Software \& Systems Modeling}, vol.~18,
  pp.~2821--2842, 2019.

\bibitem{bagherzadeh2019live}
M.~Bagherzadeh, K.~Jahed, B.~Combemale, and J.~Dingel, ``Live-umlrt: a tool for
  live modeling of uml-rt models,'' in {\em 2019 ACM/IEEE 22nd International
  Conference on Model Driven Engineering Languages and Systems Companion
  (MODELS-C)}, pp.~743--747, IEEE, 2019.

\bibitem{bagherzadeh2021live}
M.~Bagherzadeh, K.~Jahed, B.~Combemale, and J.~Dingel, ``Live modeling in the
  context of state machine models and code generation,'' {\em Software and
  Systems Modeling}, vol.~20, pp.~795--819, 2021.

\bibitem{exelmans2024operation}
J.~Exelmans, C.~Teodorov, and H.~Vangheluwe, ``Operation-based versioning as a
  foundation for live executable models,'' {\em Software and Systems Modeling},
  pp.~1--19, 2024.

\bibitem{lopez2019multiverse}
C.~T. Lopez, R.~G. Singh, S.~Marr, E.~G. Boix, and C.~Scholliers, ``Multiverse
  debugging: Non-deterministic debugging for non-deterministic programs,'' in
  {\em 33rd European Conference on Object-Oriented Programming}, vol.~134,
  pp.~1--27, 2019.

\bibitem{pasquier2022practical}
M.~Pasquier, C.~Teodorov, F.~Jouault, M.~Brun, L.~L. Roux, and L.~Lagadec,
  ``Practical multiverse debugging through user-defined reductions: Application
  to uml models,'' in {\em Proceedings of the 25th International Conference on
  Model Driven Engineering Languages and Systems}, pp.~87--97, 2022.

\bibitem{pasquier2023debugging}
M.~Pasquier, C.~Teodorov, F.~Jouault, M.~Brun, and L.~Lagadec, ``Debugging
  paxos in the uml multiverse,'' in {\em 2023 ACM/IEEE International Conference
  on Model Driven Engineering Languages and Systems Companion (MODELS-C)},
  pp.~811--820, IEEE, 2023.

\bibitem{cimino2021harmonising}
C.~Cimino, G.~Ferretti, and A.~Leva, ``Harmonising and integrating the digital
  twins multiverse: A paradigm and a toolset proposal,'' {\em Computers in
  Industry}, vol.~132, p.~103501, 2021.

\end{thebibliography}
